\documentclass{eptcs}
\usepackage[T1]{fontenc}

\title{Towards Relating \ciao Assertions and \lptp Theorems
  \thanks{%
    Partially funded by MICIU projects CEX2024-001471-M \emph{María de
      Maeztu} and TED2021-132464B-I00 \emph{PRODIGY}, 
      and by the European Union GA 101154447 NEAT. 
    We would also like to thank the anonymous reviewers for their very
    useful and constructive feedback.%
  }%
}
\author{\hspace*{-0.5em}Marco P\'erez$^{1,2}$
  \href{https://orcid.org/0000-0002-1092-2071}{Pedro Lopez-Garcia}$^{3,2}$
  \href{https://orcid.org/0000-0001-9782-8135}{Jose F. Morales}$^{1,2}$
  \href{https://orcid.org/0000-0002-7583-323X}{Manuel V. Hermenegildo}$^{1,2}$ %
  \href{https://orcid.org/0000-0002-8775-7430}{Fred Mesnard}$^{4}$
  \vspace{1mm}
  \institute{\hspace*{-0.5em}$^{1}$Universidad Polit\'ecnica de Madrid (UPM) Madrid, Spain
   \hspace{6em} $^{2}$IMDEA Software Institute, Madrid, Spain}
  \institute{$^{3}$Spanish Council for Scientific Research
    \hspace{10em} $^{4}$Universit\'e de La R\'eunion}
  \email{\hspace*{-1.5em}\{marco.perez,pedro.lopez,josef.morales,manuel.hermenegildo\}@imdea.org,
    frederic.mesnard@univ-reunion.fr}
}
\newcommand{\titlerunning}{Towards Relating \ciao Assertions and \lptp Theorems}
\newcommand{\authorrunning}{M. Perez, P. L\'opez-Garc\'ia, J.F. Morales,
  M.V. Hermenegildo, F. Mesnard}

\usepackage{stmaryrd}
\usepackage{pdflscape} 
\usepackage{afterpage}
\usepackage{graphicx}
\usepackage{subcaption}
\DeclareCaptionLabelFormat{Anumber}{1#2}
\graphicspath{ {./plots/} }
\usepackage{changepage}
\usepackage{algorithm}
\usepackage[noend]{algpseudocode}
\usepackage{adjustbox}
\usepackage{ amssymb }
\usepackage{ amsmath }
\usepackage{mathtools}
\usepackage{stackengine}
\usepackage{tablefootnote}
\usepackage{tabularx}

\usepackage[main=english]{babel}
\usepackage{epigraph}
\usepackage{xspace}
\usepackage{listings}
\usepackage{color}

\usepackage[
table,
xcdraw,
dvipsnames]{xcolor}

\usepackage{mdwmath}
\usepackage{fancybox}
\usepackage{bookmark}
\usepackage{enumitem}
\usepackage{setspace}
\usepackage[acronym]{glossaries}
\usepackage{siunitx}

\hypersetup{hidelinks,pageanchor=true,colorlinks,citecolor=Fuchsia,urlcolor=black,linkcolor=Cerulean}

\graphicspath{{./graphics/}}

\usepackage{float}

\DeclareGraphicsExtensions{.pdf,.jpeg,.jpg,.png}

\definecolor{LightCyan}{rgb}{0,0,0}
\definecolor{Cerulean}{rgb}{0,0,0}
\definecolor{Fuchsia}{rgb}{0,0,0}

\usepackage{float}
\newfloat{algorithm}{t}{lop}

\definecolor{mGreen}{rgb}{0,0.6,0}
\definecolor{mGray}{rgb}{0.5,0.5,0.5}
\definecolor{mPurple}{rgb}{0.58,0,0.82}
\definecolor{backgroundColour}{rgb}{0.95,0.95,0.92}

\usepackage{listings}

\definecolor{ciaoframe}     {rgb}{  0,    0,  0.3}
\definecolor{ciaostring}    {rgb}{0.6, 0.46, 0.33}
\definecolor{ciaooperators} {rgb}{0.1, 0.15,  0.6}
\definecolor{ciaokeywords}  {rgb}{0.1, 0.15,  0.6}
\definecolor{ciaoassertions}{rgb}{0.1, 0.15,  0.6}
\definecolor{ciaotrust}     {RGB}{200, 130,     0}
\definecolor{ciaocheck}     {rgb}{0.1, 0.2,   0.8}
\definecolor{ciaochecked}   {rgb}{0.2, 0.34,  0.1}
\definecolor{ciaotrue}      {rgb}{0.2, 0.34,  0.1}
\definecolor{ciaofalse}     {rgb}{0.6,  0.0, 0.09}
\definecolor{ciaoprops}     {rgb}{0.1,  0.2,  0.8}
\definecolor{ciaocomment}   {rgb}{0.8,  0.3,  0.3}

\newcommand{\prettylstciao}[0]{
\lstset{language=Prolog,
  frameround=fttt,
  frame=ltrb,
  rulecolor=\color{ciaoframe},
  numbers=left,numberstyle=\tiny,stepnumber=1,numbersep=8pt,
  tabsize=4,
  breaklines=true,breakatwhitespace=true,
  basicstyle=\scriptsize\ttfamily, %
  showlines=true,
  showspaces=false,
  showtabs=false,
  escapechar=@,
  escapeinside={~~},
  commentstyle=\color{ciaocomment},
  stringstyle=\color{ciaostring},
  showstringspaces=false,
  deletekeywords={true}, %
  keywordstyle={\color{ciaooperators}\bfseries}, %
  classoffset=1, %
        otherkeywords={>,<,>=,=<,.,;,-,!,=,*,\&,+,:-,[,],|,->,:,:=,\#},
        keywordstyle={\color{ciaokeywords}\bfseries},
  classoffset=2,
       morekeywords={module,use_module,dynamic,export,import,multifile,impl_defined,trait,impl,mode},
       keywordstyle={\color{ciaokeywords}\bfseries},
       morekeywords={pred,prop,calls,success,comp},
       keywordstyle={\color{ciaoassertions}\bfseries},
  classoffset=4,
       morekeywords={trust,trust_default,entry},
       keywordstyle={\color{ciaotrust}\bfseries},
  classoffset=5,
       morekeywords={check},
       keywordstyle={\color{ciaocheck}\bfseries},
  classoffset=6,
       morekeywords={checked},
       keywordstyle={\color{ciaochecked}\bfseries},
  classoffset=7,
       morekeywords={true},
       keywordstyle={\color{ciaotrue}\bfseries},
  classoffset=8,
       morekeywords={false},
       keywordstyle={\color{ciaofalse}\bfseries},
  classoffset=9,
       morekeywords={even,nat,str,int,flt,atm,term,num,var,list,ground,mshare,
                    rsize,cardinality,not_fails,exp,cost,costb,steps_ub,steps_lb,
                    size_ub,size_lb,covered,mut_exclusive,head_cost,literal_cost,
                    is_det,non_det,length,terminates,steps_o,resource,socket,seff,string,
                    lowercase
       },
       keywordstyle={\color{ciaoprops}\bfseries},
  classoffset=0, %
}}

\usepackage{tikz}
\usetikzlibrary{shapes.geometric}
\usetikzlibrary{matrix}
\usetikzlibrary{shapes.gates.logic.US}
\usetikzlibrary{circuits}
\usetikzlibrary{positioning,shapes,arrows,fit,shadows,calc,matrix}
\usetikzlibrary{backgrounds}

\usetikzlibrary{arrows}
\usetikzlibrary{arrows.meta}

\usepackage{forest}

\makeatletter
\csdef{forest@compute@node@boundary@rounded rectangle}{%
  \forest@mt{east}%
  \forest@lt{north east}%
  \forest@lt{north}%
  \forest@lt{north west}%
  \forest@lt{west}%
  \forest@lt{south west}%
  \forest@lt{south}%
  \forest@lt{south east}%
  \forest@lt{east}%
}
\makeatother
\newtheorem{definition}{Definition}%
\newtheorem{theorem}{Theorem}[section]

\usepackage{wrapfig}

\newcommand{\secbeg}{}
\newcommand{\secend}{}

\newcommand{\exabeg}{}
\newcommand{\exaend}{}

\newcommand{\tuple}[3]{\ensuremath{\tuplek{\langle #1,\allowbreak #2\rangle}{#3}\xspace}}

\DeclareOldFontCommand{\tt}{\ttfamily}{\mathtt}

\definecolor{lightred}{rgb}{0.8,  0.3,  0.3} 
\definecolor{forestgreen}{rgb}{34, 139, 34}

\usepackage{pifont}%

\renewcommand{\paragraph}[1]{\noindent\textbf{\emph{#1}}}

\usepackage{booktabs} %

\definecolor{ciaoframe}     {rgb}{  0,    0,  0.3}
\definecolor{ciaostring}    {rgb}{0.6, 0.46, 0.33}
\definecolor{ciaooperators} {rgb}{0.1, 0.15,  0.6}
\definecolor{ciaokeywords}  {rgb}{0.1, 0.15,  0.6}
\definecolor{ciaoassertions}{rgb}{0.1, 0.15,  0.6}
\definecolor{ciaotrust}     {RGB}{200, 130,     0}
\definecolor{ciaocheck}     {rgb}{0.1, 0.2,   0.8}
\definecolor{ciaochecked}   {rgb}{0.2, 0.34,  0.1}
\definecolor{ciaotrue}      {rgb}{0.2, 0.34,  0.1}
\definecolor{ciaofalse}     {rgb}{0.6,  0.0, 0.09}
\definecolor{ciaoprops}     {rgb}{0.1,  0.2,  0.8}
\definecolor{ciaocomment}   {rgb}{0.5,  0.5,  0.5}

\newcommand{\clause}{clause\xspace}

\newcommand{\prim}[1]{\ensuremath{{#1}^{\prime}}}
\newcommand{\noclause}{\epsilon}

\newcommand{\tightoverset}[2]{%
  \mathop{#2}\limits^{\vbox to -.5ex{\kern-0.75ex\hbox{{\scriptsize #1}}\vss}}}

\newcommand{\transit}[2]{\overset{#2\ }{\leadsto_{#1}}}
\newcommand{\transitstar}[2]{\tightoverset{#2\ \ }{\leadsto^{*}_{#1}}}
\newcommand{\transitp}[1]{\overset{#1}{\leadsto}}
\newcommand{\transitstarp}[1]{\tightoverset{#1\ \ }{\leadsto^{*}}}

\newcommand{\derivation}{\ensuremath{d}\xspace}
\newcommand{\derivationset}{\ensuremath{\mathit{D}}\xspace}
\newcommand{\branch}{\ensuremath{\mathcal{B}}\xspace}

\newcommand{\state}{\ensuremath{S}\xspace}
\newcommand{\program}{\ensuremath{P}\xspace}

\newcommand{\currgoal}[1]{\ensuremath{\mathit{curr\_goal}(#1)}\xspace}
\newcommand{\currstore}[1]{\ensuremath{\mathit{curr\_store}(#1)}\xspace}
\newcommand{\currstate}[1]{\ensuremath{\mathit{curr\_state}(#1)}\xspace}
\newcommand{\currclause}[1]{\ensuremath{\mathit{curr\_clause}(#1)}\xspace}
\newcommand{\clseq}[1]{\ensuremath{\mathit{clseq}(#1)}\xspace}
\newcommand{\laststate}[1]{\ensuremath{\mathit{last\_state}(#1)}\xspace}
\newcommand{\initstate}[1]{\ensuremath{\mathit{init\_state}(#1)}\xspace}

\newcommand{\successors}[1]{\ensuremath{\mathit{successors}(#1)}\xspace}
\newcommand{\leaf}[1]{\ensuremath{\mathit{leaf}(#1)}\xspace}
\newcommand{\lastderiv}[1]{\ensuremath{\mathit{last\_deriv}(#1)}\xspace}

\newcommand{\succrelation}[2]{\ensuremath{#1 \, \mathit{succ} \, #2}\xspace}
\newcommand{\intermsofderivations}[1]{}
\newcommand{\searchstr}{\ensuremath{\mathit{str}}\xspace}
\newcommand{\dfstr}{\ensuremath{\mathit{df}}\xspace}

\newcommand{\finite}{\ensuremath{\mathit{finite}}\xspace}
\newcommand{\infinite}{\ensuremath{\mathit{infinite}}\xspace}
\newcommand{\finished}{\ensuremath{\mathit{finished}}\xspace}
\newcommand{\successful}{\ensuremath{\mathit{successful}}\xspace}
\newcommand{\failed}{\ensuremath{\mathit{failed}}\xspace}
\newcommand{\terminates}{\ensuremath{\mathit{terminates}}\xspace}
\newcommand{\noterminates}{\ensuremath{\mathit{does\_not\_terminate}}\xspace}
\newcommand{\succeeds}{\ensuremath{\mathit{succeeds}}\xspace}
\newcommand{\fails}{\ensuremath{\mathit{fails}}\xspace}
\newcommand{\finitelysucceeds}{\ensuremath{\mathit{finitely\_succeeds}}\xspace}
\newcommand{\infinitelysucceeds}{\ensuremath{\mathit{infinitely\_succeeds}}\xspace}
\newcommand{\finitelyfails}{\ensuremath{\textit{finitely\_fails}}\xspace}
\newcommand{\infinitelyfails}{\ensuremath{\mathit{infinitely\_fails}}\xspace}

\newcommand{\finitelyfailstext}{\ensuremath{\mathit{finitely \; fails}}\xspace}

\providecommand{\success}{}\renewcommand{\success}{\ensuremath{\mathit{success}}\xspace}
\providecommand{\failure}{}\renewcommand{\failure}{\ensuremath{\mathit{failure}}\xspace}
\providecommand{\termination}{}\renewcommand{\termination}{\ensuremath{\mathit{termination}}\xspace}
\newcommand{\nontermination}{\ensuremath{\textit{non--termination}}\xspace}

\newcommand{\triple}[3]{\langle #1, #2, #3 \rangle} 
\newcommand{\neck}{\mbox{\tt :- }}
\def\tuple#1{\langle #1 \rangle}

\newcommand{\Q}{\mbox{$\mathit{Q}$}}
\newcommand{\project}[2]{\overline{\exists}_{#1}#2}
\newcommand{\query}{\ensuremath{Q}\xspace}
\newcommand{\queryset}{\ensuremath{\mathsf{Q}}\xspace}
\newcommand{\genprop}{\ensuremath{\mathit{prop}}\xspace}

\newcommand{\lptpS}{\ensuremath{\mathbf{S}}\;}

\newcommand{\lptpT}{\ensuremath{\mathbf{T}}\;}

\newcommand{\successAsrN}{\textsf{success}}
\newcommand{\successAsr}[3]{\ensuremath{\texttt{:- success } #1: #2 \texttt{ => } #3}}
\newcommand{\compAsrN}{\textsf{comp}}
\newcommand{\compAsr}[3]{\ensuremath{\texttt{ :- comp } #1: #2 \texttt{ + } #3}}

\newcommand{\lptp}{\texttt{LPTP}\xspace}
\newcommand{\ciao}{\texttt{Ciao}\xspace}
\newcommand{\ciaoprolog}{\texttt{Ciao Prolog}\xspace}
\newcommand{\ciaopp}{\texttt{CiaoPP}\xspace}

\newcommand{\prolog}{\texttt{Prolog}\xspace}

\newcommand{\pre}{\ensuremath{\mathit{Pre}}\xspace}
\newcommand{\post}{\ensuremath{\mathit{Post}}\xspace}
\newcommand{\comprops}{\ensuremath{\mathit{Comp}}\xspace}
\newcommand{\status}{\ensuremath{\mathit{Status}}\xspace}

\newcommand{\achecked}{\textcolor{ciaochecked}{checked}}
\newcommand{\acheck}{\textcolor{ciaocheck}{check}}
\newcommand{\afalse}{\textcolor{ciaofalse}{false}}
\newcommand{\atrue}{\textcolor{ciaotrue}{true}}
\newcommand{\M}{M}

\newcommand{\vv}{{\bar v}}
\newcommand{\pv}{p(\bar v)}
\newcommand{\pvp}{p(\prim{\bar v})}

\newcommand{\lsem}{\mbox{$\llbracket$}}
\newcommand{\rsem}{\mbox{$\rrbracket$}}
\newcommand{\sem}[1]{\lsem #1 \rsem}

\newcommand{\psem}{\sem{\program}_{\Q}}

\newcommand{\analysis}{\ensuremath{\psem^\alpha}}
\newcommand{\renanalysis}[1]{\ensuremath{{\it Ren}(#1, \analysis)}}

\begin{document}

\maketitle

\begin{abstract}
Abstract interpretation–based verification is a central component of
the \ciaoprolog system, enabling expressive specifications of
properties of programs, predicates, and execution
states. Independently, the \lptp (Logic Programming Theorem Proving)
framework offers a first-order logical formalism for expressing and
proving properties of predicates.  In this paper, we address a
fundamental issue in relating these two frameworks: studying the
translation of \ciao assertions into \lptp formulae and 
identifying a partial correspondence between assertion-based and
logic-based specifications.
We introduce a systematic translation scheme, characterize assertion
classes according to their logical encodability, and propose
approximation strategies and auxiliary constructs for non-translatable
cases, and finally analyze the resulting soundness and completeness trade-offs.
We 
argue that our proposal enables a tight integration
of \ciao’s assertion checking with \lptp-based deductive verification, thereby
leveraging their complementary capabilities.
\end{abstract}
	
\secbeg
\section{Introduction}
\secend

The landscape of formal verification is defined by a fundamental
trade-off between automation and generality. Abstract interpretation
offers a powerful framework for fully automated analysis: by reasoning
over %
abstract domains, it allows for ``push-button''
verification of program properties. However, this automation comes at
the cost of precision and expressiveness; properties must be
approximated, and complex logical relationships may be lost in the
abstraction. Conversely, interactive theorem proving offers maximum
generality and precision, capable of reasoning about arbitrarily
complex mathematical properties. Yet, this power requires significant
manual intervention, making it labor-intensive and difficult to scale
to large codebases.

In the context of logic programming, \ciao's assertion framework and
the \lptp (Logic Programming Theorem Proving) system serve as clear
representatives of these respective paradigms. Abstract
interpretation--based verification is a central component of the
\ciao system's preprocessor, \ciaopp~\cite{aadebug97-informal,prog-glob-an,assrt-theoret-framework-lopstr99,ciaopp-sas03-journal-scp,intermod-incanal-2020-tplp},
enabling expressive specifications of properties
of programs,
predicates, and execution states. Within this framework, assertions
serve as machine-understandable specifications that support program
validation, debugging, and optimization through static
analysis. %
In turn, the \lptp
framework~\cite{LPTP-Stark97,LPTP-jlp-Stark98,LPTP-manual} offers a
logical formalism 
aimed at representing logic programs and their properties as
first-order formulae suitable for interactive theorem proving, also
suitable for partial automation using modern proof
systems~\cite{mesnard-auto-tp-prolog-verification}.
While both systems target the verification of logic programs, they do
so from complementary perspectives: \ciao assertions
are rooted in the approximation of execution states, and can thus
cover operational aspects, whereas \lptp is centered on the logical
aspects for which it offers
proof-theoretic generality. Consequently, it makes compelling sense to
combine these distinct paradigms within a single verification
platform. A hybrid system can leverage abstract interpretation to
automatically discharge simple proof obligations and infer invariants,
while reserving the heavy lifting of theorem proving for complex
properties that escape static analysis.

In this paper, we address the problem of relating these two frameworks
by studying the translation of \ciao assertions into \lptp formulae. We
aim to enable a tight integration of \ciao's assertion checking with
\lptp-based deductive verification. Our contributions are as follows:
Based on some background and preliminary information 
we develop in 
Sections~\ref{sec:background}--\ref{sec:lptpsemantics} our formalization.
We then identify
in Section~\ref{sec:sldsemanticsandlptp} a partial correspondence
between assertion-based and logic-based specifications, showing that
while many assertions translate faithfully, others resist direct
encoding due to intrinsic semantic mismatches.  We then introduce in
Sections \ref{sec:ciao_to_logic} and \ref{sec:prop_corresp} a partial
translation scheme and characterize assertion classes according to
their logical encodability. We finally propose in
Section~\ref{sec:prop_corresp} approximation strategies and auxiliary
constructs for those assertions that fall outside the direct
expressive scope of first-order logic, and we conclude in
Section~\ref{sec:conclusions}.

\secbeg
\section{Background}
\label{sec:background}
\secend

\paragraph{\ciao Assertions.}
\ciao assertions act as a bridge between the program text and the
formal verification machinery of the \ciao Preprocessor
(\ciaopp)~\cite{aadebug97-informal,prog-glob-an}.
Assertions are first-class
syntactic entities that can be processed statically, dynamically, or
effectively ignored during standard execution. A primary assertion
unit is the \texttt{pred} declaration, which simultaneously describes
call patterns, success states, and computational behavior:
\vspace*{-3mm}
\[
\texttt{:-} \ \mbox{\status}  \texttt{ pred } p(X_1, \dots, X_n) : \mathit{Pre} \Rightarrow \mathit{Post} + \mathit{Comp}.
\vspace*{-2.5mm}
\]
which is syntactic sugar for three distinct assertion types that
isolate different operational
phases~\cite{assrt-theoret-framework-lopstr99,ciaopp-sas03-journal-scp}:

\begin{itemize}
  \itemsep=-3pt
    \item \textbf{Calls Assertions (\texttt{:- calls}):} The
      $\mathit{Pre}$ field specifies properties that must hold
      \emph{at the moment the predicate is invoked}. This is
      inherently operational, constraining the set of admissible
      execution states (e.g., $\texttt{ground(X)}$,
      $\texttt{list(L)}$).

    \item \textbf{Success Assertions (\texttt{:- success}):} The
      $\mathit{Post}$ field describes the properties that must hold in
      the success state. It is a logical implication: \emph{if} the
      precondition holds at the call, \emph{then} the postcondition
      must hold upon success.

    \item \textbf{Comp Assertions (\texttt{:- comp}):} The
      $\mathit{Comp}$ field captures global computational properties
      of the execution, such as determinism, termination, cost, or
      non-failure.
\end{itemize}
\vspace*{-1mm}

\noindent
$\status$ is a qualifier of the meaning of the assertion:
\vspace*{-1 mm}
\begin{itemize}
\itemsep=-2pt
\item {\tt \atrue}: the assertion expresses properties inferred by
  static analysis (correct over-approximation of the concrete semantic
  properties).

\item {\tt \acheck}: the assertion expresses properties that must hold
  at run-time, i.e., that the analyzer should prove (or else generate
  run-time checks for). This is the \emph{default} status, and can be
  omitted.
\item {\tt \achecked}: the analyzer proved that the property holds in all executions.

\item {\tt \afalse}: the analyzer proved that the property does not
  hold in some execution.

\item \texttt{trust}: provides the analyzer with information that it
  should \emph{assume}. This can be useful for, e.g., describing
  external or unknown code or improving analysis precision.

\end{itemize}

\ciaopp generally proves or disproves assertions by safely
approximating program properties through abstract interpretation-based
static analysis (we return to this issue in
Section~\ref{sec:assertion-checking}). 

\medskip
\paragraph{\lptp Formulae and the Inductive Extension.}
The \lptp (Logic Programming Theorem Proving) system operates on a
different theoretical basis, treating logic programs not %
as instructions for a resolution engine but as mathematical objects in an
\emph{inductive extension} of first-order
logic~\cite{LPTP-Stark97,LPTP-jlp-Stark98}. This framework transforms
a \prolog program $P$ into a formal theory $\mathit{IND}(P)$ that
includes Clark's completion and specific induction schemas for each
predicate.
Properties in \lptp are expressed using three primary logical
operators $\textbf{S}$, $\textbf{F}$, $\textbf{T}$ that model the
operational semantics of success, failure and termination
respectively.
Unlike \ciao assertions, which, as mentioned before, are typically checked
using abstract domains (lattices of approximations), \lptp formulae
are verified via interactive proof using a Gentzen-style sequent
calculus~\cite{LPTP-manual}. This allows \lptp to prove complex
properties that require the development of specific domains in
abstract interpretation.

\secbeg
\section{SLD Semantics}
\label{sec:sldsemantics}
\secend

We will denote by ${\cal C}$ the universal set of constraints.
We let $\project{L}{\theta}$ be the constraint $\theta$ restricted to
the variables of the syntactic object $L$.  We denote constraint
entailment by $\models$, so that $c_1\models c_2$ denotes that $c_1$
entails $c_2$.

An \emph{atom} has the form $p(t_1,...,t_n)$ where $p$ is a predicate
symbol and the $t_i$ are terms.  A \emph{literal} is either an atom
or a constraint.  A \emph{goal} is a finite sequence of
literals.  A \emph{rule} is of the form $H \neck B$ where $H$,
the \emph{head}, is an atom and
$B$, the \emph{body}.
A fact is a rule whose \emph{body} is the literal \texttt{true}.

A \emph{constraint logic program}, or \emph{program}, is a finite set
of rules.  The \emph{definition} of an atom $A$ in program $\program$,
${\mathit defn}_P(A)$, is the set of variable renamings of rules in
$\program$ such that each renaming has $A$ as a head and has distinct new
local (but not head) variables.
We assume that every rule in the program has an associated unique
identifier. Rule identifiers are denoted by $c$, usually subscripted.

The operational semantics of a program is in terms of its
``derivations'' which are sequences of reductions between ``states''.
A \emph{state} $\triple{G}{\theta}{c}$ consists of a goal $G$, a
constraint store (or \emph{store} for short) $\theta$, and a clause
identifier $c$.  A state $\triple{L::G}{\theta}{\_}$,
where $L$ is a
literal and $::$ denotes concatenation of sequences, can be
\emph{reduced} as follows:
if $L$ is a constraint and $\theta \land L$ is satisfiable, it
   is reduced to $\triple{G}{\theta \land L}{\noclause}$.  Note that
   \texttt{true} is considered a constraint (builtin) such that
   $\theta \land \text{ true } \equiv \theta$; 
   if $L$ is an atom, it is reduced to
   $\triple{B::G}{\theta}{c}$ for some rule $(L \mbox{\tt :-} B)$
   $\in defn_P(L)$ whose identifier is $c$.

We assume for simplicity that the underlying constraint solver is
complete.
We use $\state \transit{\program}{c} \prim{\state}$ to indicate that in program $\program$ a
reduction can be applied to state $\state$ using clause $c$ to obtain state
$\prim{\state}$. Also, $\state \transitstar{\program}{C} \prim{\state}$ indicates that there is a
sequence of reduction steps from state $\state$ to state $\prim{\state}$ using the
sequence of clauses $C$. When program $\program$ is understood from the context, we write
$\state \transitp{c} \prim{\state}$ and $S \transitstarp{C} \prim{\state}$.
When we want to express
$S \transitp{c} \prim{\state}$ or $S \transitstarp{C} \prim{\state}$
for some $c$ or $C$ that do not need
to specify, we write
$S \transitp{ } \prim{\state}$ or $S \transitstarp{ } \prim{\state}$
respectively.

A \emph{derivation} from state $\state$ for program $\program$ is a finite
sequence of states $\state_0, \state_1, \ldots, \state_n$, $n \geq 0$,
where $\state_0$ is $\state$, each $\state_i$ is of the form
$\triple{G_i}{\theta_i}{c_i}$, and for each $1 \leq i \leq n$,
$\state_{i-1} \transit{\program}{c_i} \state_i$. I.e., the derivation represents the
sequence of reductions $S \transit{\program}{c_1} \state_1 \cdots \state_{i-1}
\transit{\program}{c_{i}} \state_i \cdots \transit{\program}{c_n} \state_n$.  Given a
non-empty derivation \derivation, its initial and last state are denoted by
$\initstate{\derivation}$ and $\laststate{\derivation}$ respectively.  The sequence of
clauses used in a derivation is denoted by $\clseq{\derivation}$.  For example,
if \derivation is the derivation above, then $\clseq{\derivation} = c_1 \cdot c_2 \cdots
c_n$.  Note that a derivation \derivation can be characterized by $\clseq{\derivation}$
and $\initstate{\derivation}$. In addition, $\currgoal{\derivation}$ and
$\currstore{\derivation}$ denote the first subgoal
and the store in $\laststate{\derivation}$, respectively.  E.g., if
$\state_n =\triple{g::G}{\theta}{c}$,
$\currgoal{\derivation}=g$ and $\currstore{\derivation}=\theta$. We also denote by
$\currclause{\derivation}$ the clause used to derive $\state_n$, i.e., $c$.
The set of all derivations from a state $\state$ for program $\program$ is
represented by function $derivations(\program,\state)$. When $\program$ is clear
from the context, we use $derivations(\state)$ instead for simplicity.
A \emph{query} is a pair $(L,\theta)$ where $L$ is a goal
and $\theta$ a store.  We extend the $derivations(\program,\state)$
function to queries, by associating the initial state
$\triple{L}{\theta}{\noclause}$ to query $(L,\theta)$ as follows:
\vspace*{-3mm}
\begin{equation}
derivations(\program,I) =
    \begin{cases*}
      derivations(\program, \triple{L}{\theta}{\noclause}) & if $I$ is a query $(L,\theta)$ \\[-2pt]
      derivations(\program,\state)  & if $I$ is a state $\state$ \\[-2pt]
    \end{cases*}
\end{equation}    
\vspace*{-3mm}

We further extend the $derivations$ function to deal with sets of
\emph{queries}, denoted by $\cal{Q}$, as follows:
$derivations(\program,{\cal{Q}})=\bigcup_{Q\in{\cal Q}} derivations(\program,\query)$.
Note that $derivations(\program,I)$ can be infinite, but any derivation $d
\in derivations(\program,I)$ is always a finite sequence of states.

A derivation \derivation is \emph{finished} iff $\laststate{\derivation}$ cannot be reduced.
Note also that $derivations(\program, I)$ contains not only finished
derivations but also all intermediate derivations.
A finished derivation $\derivation$ is \emph{successful} iff $\laststate{\derivation}$ is of the
form $\triple{nil}{\theta}{c}$, where $nil$ denotes the empty
sequence.
A finished derivation is \emph{failed} iff $\laststate{\derivation}$ is not of the
form $\triple{nil}{\prim{\theta}}{c}$.

\exabeg
\begin{definition}[Successor relation between derivations]
Given a derivation $\derivation = \state_0, \state_1, \ldots, \state_n$, $n \geq 0$, we say
that derivation $\prim{\derivation}$ is a successor of it, denoted by $\succrelation{\derivation}{\prim{\derivation}}$, iff $\prim{\derivation}$
is of the form $\derivation = \state_0, \state_1, \ldots, \state_n, \state_{n+1}$ and there is a
\clause $c_{n+1}$ such that $\state_n \transit{\program}{c_{n+1}} \state_{n+1}$.  We
denote by $\successors{\derivation}$ the set of all successors of derivation $\derivation$,
i.e., $\successors{\derivation} = \{ \prim{\derivation} \,|\, \succrelation{\derivation}{\prim{\derivation}} \}$.
\end{definition}
\exaend

\paragraph{\large{\textbf{Derivations and Trees}}.}
Given state $\state$, we represent $derivations(\program,\state)$ as an
SLD-tree, denoted by $tree(derivations(\program,\state))$ or simply
$tree(\program,\state)$, whose root is $\state$.

\exabeg
\begin{definition}[Branch]
A branch is a (possibly infinite) subset $\branch \subseteq
derivations(\program,\state)$ such that all the derivations in $\branch$ can be arranged
as a sequence $\derivation_0, \ldots, \derivation_{i-1}, d_i, \ldots$
such that
for all $i \geq 1$, $\derivation_i \in successors(\derivation_{i-1})$.
\end{definition}

\noindent
Note that a branch $\branch$ can be either \emph{finite} or
\emph{infinite}, depending on whether the set $\branch$ has a
\emph{finite} or \emph{infinite} number of elements, respectively.

Given a \emph{finite} branch $\branch$, ordered as $\derivation_0, \derivation_1, \ldots, \derivation_n$
according to the successor relation, we denote its last derivation as
$\lastderiv{\branch} = \derivation_n$.  We sometimes represent $\branch$ as just the
derivation $\derivation_n$ by abuse of notation, since the rest of derivations
$\derivation_0, \derivation_1, \ldots, \derivation_{n-1}$ is redundant and can be obtained from
$\derivation_n$.

\exabeg
\begin{definition}[Leaf of a finite branch]
Given a \emph{finite} branch $\branch$,
we define its leaf as $\leaf{\branch} = \laststate{\lastderiv{\branch}}$.
\end{definition}
\exaend

\exabeg
\begin{definition}[Tree of a set of derivations]
We define $tree(\program,\state) = \{ \branch \in derivations(\program,\state) \,|\, $\branch$ \text{ is a
  branch} \}$
\end{definition}
\exaend

In the graphical representation of a tree, the prefixes that are
common to several branches are depicted together.  The leaf of a
branch $\branch$ is expanded with a sibling, each one per state $S \in
\{ \laststate{A} \mid A \in \successors{\lastderiv{\branch}}\}$.

Let $\branch$ be a branch. We say that $\branch$ is \emph{finished} iff it is \emph{finite} and
$\lastderiv{\branch}$ is \emph{finished};
$\branch$ is \emph{successful} iff it is \emph{finished} and
$\lastderiv{\branch}$ is \emph{successful};
$\branch$ is \emph{failed} iff it is \emph{finished} and
$\lastderiv{\branch}$ is \emph{failed}.

\secbeg
\subsection{Properties of Queries Referred to SLD-Trees} 
\label{sec:prop-trees-breadth}
\secend

We now express the properties of a query $\query$
in terms of its complete SLD-Tree, i.e., the one resulting from the
breadth-first evaluation of $\query$ (also referred to as the resolution
tree). We will call these properties ``universal.''

Given a query $\query$ of the form $(L,\theta)$ for program $\program$, if there
is a \successful branch $\branch \in tree(\program,\query)$ such that $\leaf{\branch}$ is
of the form $\triple{nil}{\prim{\theta}}{c}$, then we say that the
constraint $\project{L}{\prim{\theta}}$ is an \emph{answer} to $\query$. We
denote by $answers(\program,\query)$ the set of all answers to query $\query$.

A query $\query$ \emph{does not terminate} for $\program$, denoted
$\noterminates(\program,\query)$, iff $tree(\program,\query)$ has at least one \infinite
branch.

\intermsofderivations{$derivations(\program,\query)$ is infinite.}

Note that the condition ``$tree(\program,\query)$ has at least one
infinite branch'' is equivalent to ``$tree(\program,\query)$ is
infinite or $derivations(\program,\query)$ is infinite.'' It also
includes the case where the tree has an infinite number of branches.%
\footnote{Note that that if there is an infinite number of finite
  branches, then necessarily there is at least one infinite
  branch. For example, if we call \texttt{nat(X)} with the standard
  definition, the branch that always chooses the recursive clause is
  infinite.}

A query $\query$ \terminates for $\program$, denoted by $\terminates(\program,\query)$, iff
$tree(\program,\query)$ does not have any infinite branch. It is the negation of
\noterminates:
\vspace*{-3mm}
$$\terminates(\program,\query) \leftrightarrow \neg \noterminates(\program,\query)$$\\ [-7mm]
Hence, $\noterminates(\program,\query) \leftrightarrow \neg \terminates(\program,\query)$.

\intermsofderivations{$derivations(\program,\query)$ is finite.}

A query $\query$ \emph{succeeds} for $\program$, denoted by $\succeeds(\program, \query)$, iff
$tree(\program,\query)$ has one \successful branch at least.  Note that any other
branch can be either \successful, \failed, or \infinite.

\intermsofderivations{$derivations(\program,\query)$ contains at least one \successful derivation.}

A query $\query$ \emph{fails} for $\program$, denoted by $\fails(\program, \query)$, iff
$tree(\program,\query)$ has no \successful branch. In other words, any branch is
either \failed or \infinite. It is the negation of \succeeds:
\vspace*{-3mm}
$$\fails(\program,\query) \leftrightarrow \neg \succeeds(\program,\query)$$\\ [-12 mm]
\intermsofderivations{$derivations(\program,\query)$ contains no \successful derivation.}

Note that both properties above, \succeeds and \fails, are orthogonal
to the \emph{termination} property. Thus, each of them include both
cases,
\emph{terminates} and \emph{does not terminate}. We
can differentiate these two cases as follows:

A query $\query$ \emph{finitely succeeds} for $\program$, denoted
$\finitelysucceeds(\program, \query)$, iff $tree(\program,\query)$ has at least one
\successful branch, and has no \infinite branches.  Note that any
branch that is not \successful must be \failed.
\intermsofderivations{$derivations(\program,\query)$ is finite and contains at
  least one \successful derivation.}
It can be defined as: $\finitelysucceeds(\program, \query) \leftrightarrow
\succeeds(\program, \query) \land \terminates(\program,\query)$.

A query $\query$ \emph{infinitely succeeds} for $\program$, denoted
$\infinitelysucceeds(\program,\query)$, iff $tree(\program,\query)$ has at least one
\successful branch, and at least one \infinite branch. Note that any
other branch can
be either \successful, \failed, or \infinite.
\intermsofderivations{$derivations(\program,\query)$ is \infinite and contains at
  least one \successful derivation.}
It can be defined as: $\infinitelysucceeds(\program,\query) \leftrightarrow
succeeds(\program,\query) \land \neg \terminates(\program,\query)$.

A query $\query$ \emph{finitely fails} in $\program$, denoted
$\finitelyfails(\program,\query)$, iff all branches of $tree(\program,\query)$ are \failed
(note that it implies termination).  Equivalently, $tree(\program,\query)$ is
\finite and has no \successful branch.
\intermsofderivations{$derivations(\program,\query)$ is \finite and contains no
  \successful derivation. }
It can be defined as follows: $\finitelyfails(\program,\query) \leftrightarrow
\fails(\program,\query) \land \terminates(\program,\query) \leftrightarrow \neg \succeeds(\program,\query)
\land \terminates(\program,\query)$.

A query $\query$ \emph{infinitely fails} in $\program$, denoted
$\infinitelyfails(\program,\query)$, iff $tree(\program,\query)$ has at least one \infinite
branch, and has no \successful branch.  I.e., any other branch is
either \failed or \infinite. Equivalently, $tree(\program,\query)$ is \infinite
and has no \successful branch.
\intermsofderivations{$derivations(\program,\query)$ is \infinite and contains no
  \successful derivation.}
It can be defined as follows: \\
$\infinitelyfails(\program,\query) \leftrightarrow \fails(\program,\query) \land \neg \terminates(\program,\query)
\leftrightarrow \neg \succeeds(\program,\query) \land \neg \terminates(\program,\query)$.

\medskip
\paragraph{\large{\textbf{Extending the definitions to sets of queries}}.}
We say that property \genprop holds for a set of queries \queryset and
program $\program$, denoted by \genprop(\program,\queryset), iff $\forall
\, \query \in \queryset \, : \genprop(\program,\query)$.
For example, \genprop can be a property in the set of properties $\{\terminates, \noterminates,$ 
$\succeeds, \fails, \finitelysucceeds, \infinitelysucceeds, \finitelyfails, \infinitelyfails \}$.

\secbeg
\subsection{Defining the Properties of Queries w.r.t.\ a Search Strategy}
\label{sec:prop-trees-depth}
\secend

Each search strategy generates the derivations in a given order, which
affects the observable behavior of the evaluation of a query.

Given a query $\query$ and a program $\program$, consider the set
$derivations(\program,\query)$ as defined in Section~\ref{sec:sldsemantics} for the
breadth-first search strategy (complete SLD-tree).  Let \searchstr
denote any search strategy, and \dfstr the depth-first search
strategy.
We denote by $derivations_{\searchstr}(\program,\query)$ the
set of derivations generated by search strategy $\searchstr$. Such set
can be defined in the same way as described in
Section~\ref{sec:sldsemantics} for
$derivations(\program,\query)$,
but using particular search and selection rules. In general,
$derivations_{\searchstr}(\program,\query) \subseteq derivations(\program,\query)$, which can
make the observable behavior of the evaluation of a query different
from that of breadth-first.
The order in which the derivations in
$derivations_{\searchstr}(\program,\query)$ are generated can also
change the behavior of the program w.r.t.\ other properties, such as
the order in which the solutions are produced. However we focus here
on the properties described in Sections~\ref{sec:prop-trees-breadth} 
and~\ref{sec:prop-trees-depth}.

The definitions of such properties w.r.t.\ a search strategy
$\searchstr$ are the same that the definitions already given
w.r.t.\ the breadth-first strategy, but referred to
$derivations_{\searchstr}(\program,\query)$ instead of $derivations(\program,\query)$. I.e.,
the only difference in the definitions is that we replace
$derivations(\program,\query)$ with $derivations_{\searchstr}(\program,\query)$.  For example,
according to Section~\ref{sec:sldsemantics},
$tree(\program,\state) = tree(derivations(\program,\state))$, so that
we define $tree_{\searchstr}(\program,\state) =
tree(derivations_{\searchstr}(\program,\state))$, and do similarly with
the rest of definitions, which are exactly the same: \emph{branch};
\emph{leaf}; \finite, \finished, \successful and \failed branch;
properties about queries (\terminates, \noterminates, \succeeds,
\fails, \finitelysucceeds, \infinitelysucceeds, \finitelyfails,
\infinitelyfails), etc.

\secbeg
\section{Meaning of the Static Analysis Information}
\secend

We now explain the meaning of the static analysis results, whose
representation as assertions with status {\tt \atrue} was introduced
in Section~\ref{sec:background}.
For some abstract domains, \ciaopp's static analysis infers properties
that refer to
the stores of the derivations (e.g., types/shapes, sharing, freeness,
groundness, etc.). We refer to such properties as
\emph{state properties}. For other analysis domains, the analysis
infers properties that refer to
(sub)sets of whole derivations, that we call \emph{computational} or
\emph{global} properties (e.g., success, failure, termination,
determinism, (type) covering, mutual exclusion, cost, etc.).

Given a set of queries $\queryset$ for a program $\program$, the
analysis results (abstract semantics), denoted by $\analysis$, can
conceptually be seen as a set of triples:
$\{\tuple{L_1,\lambda_{1}^c,\lambda_{1}^s}, \ldots,
\tuple{L_n,\lambda_{n}^c,\lambda_{n}^s}\}$. In each
$\tuple{L_i,\lambda_{i}^c,\lambda_{i}^s}$ triple, $L_i$ is a predicate
descriptor (an atom whose arguments are distinct variables), and
$\lambda_{i}^c$ and $\lambda_{i}^s$ are, respectively, the abstract
call and success substitutions, elements of an abstract domain
$D_\alpha$. Each pair $(L_i,\lambda_{i}^c)$ defines a calling pattern,
which corresponds to a node in the graph constructed by the analysis,
and represents a set of concrete queries. There is one tuple per
calling pattern.  We also consider tuples of the form
$\tuple{L_i,\lambda_{i}^c,\lambda_{i}^s, \beta}$, where $\beta$
represents global computational properties. However, for simplicity,
we assume that $\beta$ is represented in $\lambda_{i}^s$, which
includes both success-state and computational properties.

For each predicate $p$ in a program $P$, the predicate descriptor $\pv$
denotes a representative of the class of all normalized atoms for $p$.
Let $\renanalysis{\pv}$ denote the set of tuples resulting from the
static analysis for predicate $p \in \program$ w.r.t.\ $\queryset$ renamed
to variables $\vv$, i.e.: 
$\renanalysis{\pv} = \{\tuple{\pv,
  \lambda^{c}\sigma,\lambda^{s}\sigma} \mid \exists
\tuple{\pvp,\lambda^{c},\lambda^{s}} \in \analysis \text{ and } \sigma
\text{ is a renaming } \text{substitution s.t. } \pv=\pvp\sigma\}$.
This renaming is useful, for example, for comparison with information
in (check) assertions, among other tasks.

To this end, before presenting correctness results of the static
analysis by abstract interpretation, we provide instrumental
definitions and notation.

\exabeg
\begin{definition}[Calling Context]
Consider a program $P$, a
predicate $p$ and a set of queries \queryset.  The {\em calling
  context} of $p$ for $P$ and \queryset~ is: \\
$C(p,P,\queryset) = \{
\project{\pv}{\theta} \mid \exists \derivation \in derivations(P,\queryset) \; (\exists G \;
\currstate{\derivation} = \triple{\pv::G}{\theta}{\_} )
\}$.
\end{definition}
\exaend

\exabeg
\begin{definition}[Success Context for a Store]
Consider a program $P$, a predicate $p$, a constraint store $\theta$,
and a set of queries \queryset.  The {\em success context} of $p$ and
$\theta$ for $P$ and \queryset~ is: \\
$S(p,\theta,P,\queryset) =
\{ \project{\pv}{\prim{\theta}} \mid \exists \derivation \in derivations(P,\queryset) \;
\exists G \; ( \triple{\pv::G}{\theta}{\_} \in \derivation \land \currstate{\derivation} =\triple{G}{\prim{\theta}}{\_} ) \}$.
\end{definition}
\exaend
  
\exabeg
\begin{definition}[Success Context]
Consider a program $P$, a predicate $p$, and a set of queries \queryset.  The
{\em success context} of $p$ for $P$ and \queryset~ is: \\
$S(p,P,\queryset) = \bigcup_{\theta \in C(p,P,\queryset)} S(p,\theta,P,\queryset)$.
\end{definition}
\exaend

We can restrict the constraints in the calling and success contexts to
the variables in $\pv$ since this does not affect the evaluation of calls
and success assertions.

\exabeg
\begin{definition}[Abstract Calling Context]
  Consider a program $P$, a
predicate $p$ and a set of queries \queryset.  The {\em abstract calling
  context} of $p$ for $P$ and \queryset~ is $C^\alpha(p,P,\queryset)=$ $\{
\lambda^c \mid \exists \lambda^s \; (\tuple{\pv,\lambda^c,\lambda^s} \in \renanalysis{\pv}) \}$
\end{definition}
\exaend

Correctness of abstract interpretation guarantees that:
\vspace*{-1mm}
\begin{enumerate}
  \itemsep=-2pt
\item $\forall \tuple{\pv,\lambda^c,\lambda^s} \in \renanalysis{\pv} \;
  (\bigcup_{\theta\in\gamma(\lambda^c)}S(p,\theta,P,\queryset)) \subseteq
  \gamma(\lambda^s)$.

\item $C(p,P,\queryset) \subseteq \gamma(C^\alpha(p,P,\queryset))$.

\end{enumerate}
\vspace*{-1mm}
    
In order to ensure correctness of compile-time checking for a set of
queries \queryset, the analyzer must be provided with a suitable $\queryset_\alpha$
such that $\queryset \subseteq \gamma(\queryset_\alpha)$. In \ciaopp,
$\queryset_\alpha$ is expressed by means of {\em entry}
assertions. %

\secbeg
\section{Assertion Checking in \ciaopp}
\label{sec:assertion-checking}
\secend

We start by giving some instrumental definitions and notation. First,
$subderivation(\derivation^s, \derivation, p)$ expresses that
$\derivation^s$ is a subderivation of derivation $\derivation$ for
predicate $p$. More precisely, $\exists \derivation_1, \state,
\derivation_2$ s.t.:
\vspace*{-1mm}
\begin{itemize}
  \itemsep=-3pt
\item $\derivation = \derivation_1::\state::\derivation_2$, where $\state = \triple{\pv::G}{\theta}{\_}$.  
\item \ \ %
\vspace*{-2mm}
\begin{equation*}
\derivation^t  =
  \begin{cases*}
    \state::\derivation_3::\state_2 & \text{ if } $\derivation_2 = \derivation_3::\state_2::\derivation_4 \land \state_2 = \triple{G}{\theta_2}{\_}$
    \\[-2pt]
    & \text{ and there is no state in } $\derivation_3$ \text{ of the form } $\triple{G}{\_}{\_}$
    \\[-2pt]
  \state::\derivation_2 & otherwise \\[-2pt]
  \end{cases*}
\end{equation*}
\vspace*{-2mm}
\item $\derivation^s$ is the result of replacing any state of the form
  $\triple{G^t::G}{\theta^t}{\_}$ in $\derivation^t$ by $\triple{G^t}{
  \theta^s}{\_}$, where $\theta^s$ is $\theta^t$ projected over the
  variables of $G^t$.
\end{itemize}

If $\derivation^s$ is a subderivation of $\derivation$ for $p$, then
$\derivation^s$ can be obtained by applying the transitions
corresponding to the sequence of clauses of $\derivation^s$ to the
first state of $\derivation^s$. Given that $\derivation$ can be
represented as a sequence of AND-trees (or an AND-tree), $\derivation^s$
is a sequence of the corresponding AND-subtrees (or an AND-subtree)
whose root has predicate $p$.

\noindent
$holdscomp(\pv, \comprops, \derivationset)$ expresses that formula
$\comprops$ holds for the derivation set $\derivationset$.
$r$ and $r(O)$
represent a variable renaming and the result of applying it to some
syntactic object $O$, respectively.

\exabeg
\begin{definition}[Checked Comp]
\label{def:checked-comp-assrt}
A $\compAsrN$ assertion
$\compAsr{\pv}{\pre}{\comprops}$ is \emph{checked}
for a predicate $p \in P$ and query set $\queryset$ iff $holdscomp(\pv,\comprops, \derivationset)$,
\noindent
where $\derivationset = \{\derivation \in subderivations(P,\queryset,p) \mid
\text{ the first state of } \derivation \text{ meets } \pre, \text{ i.e., }
\derivation = \state::\derivation_1 \land \state = \triple{q::G}{\theta}{\_} \land q=r(\pv) \land \theta \models_{\M} r(Pre)\}$
\end{definition}

\exabeg
\begin{definition}[Checked Succ]
\label{def:checked-succ-assrt}
A $\successAsrN$ assertion
$\successAsr{\pv}{\pre}{\post}$ is
\emph{checked} for a predicate $p \in P$ and query set $\queryset$ iff
for all successful $\derivation \in subderivations(P,\queryset,p)$, if
the first state of $\derivation$ meets $\pre$, then the last state of
$\derivation$ meets $\post$.
\end{definition}
\exaend

\paragraph{True Assertions.}
For $\successAsrN$ and $\compAsrN$ assertions, we also define the
concept of \emph{true} assertion (in addition to \emph{checked} and
\emph{false}).  We say that an assertion is \emph{true} iff it is
\emph{checked} for every set of queries $\queryset$.
Note that an assertion condition $calls(p,\pre)$ can never be found to
be \emph{true}, as the calling context of $p$ depends on the query. If
we pose no restriction on the queries we can always find a calling
state which violates the assertion, unless $\pre$ is a tautology.
Clearly, the concept of \emph{true} assertions is strictly stronger
than that of \emph{checked} assertions. In general, an assertion that
is \emph{checked} (for a given set of queries) is not necessarily
\emph{true}.

\paragraph{Context-Independence.}
Given a $\successAsrN$ assertion $\successAsr{\pv}{\pre}{\post}$ or a
$\compAsrN$ assertion $\compAsr{\pv}{\pre}{\comprops}$,
the \emph{context-independent} set of queries of it is:
$Q((\pv, \pre)) = \{ (\pv, \theta) \mid  \theta \models_M  \pre \}$.
I.e., the set of all queries that meet precondition $\pre$.

We say that a $\successAsrN$ assertion
(resp. $\compAsrN$
assertion)
is \emph{context-independent checked} for predicate $p \in P$ iff it
is \emph{checked} for $p$ and its corresponding
\emph{context-independent} set of queries.
The concepts of \emph{context-independent checked} and \emph{true}
assertions are equivalent.

\secbeg
\section{\lptp Semantics} \label{sec:lptpsemantics}
\secend

To compare \lptp semantics with the general \prolog semantics
described above, it is important to take into account that \lptp
distinguishes two separate sets of notions of \success, \failure, and
\termination. The first set is semantic in nature and is defined in
terms of derivations, leading to concepts equivalent to those examined
previously in Section~\ref{sec:sldsemantics}.
The names used to refer to these properties in \lptp
papers and documents coincide with those used in
Section~\ref{sec:sldsemantics}, with the exception of the name
$\mathit{fails}$, which corresponds to the property called
\finitelyfailstext (or \finitelyfails) in
Section~\ref{sec:sldsemantics}. To avoid confusion and to adopt a unified
and as standard as possible nomenclature, in this paper we use the
name \finitelyfailstext instead of the \lptp name $\mathit{fails}$
when referring to this property, including in the context of \lptp.
We reserve the name \fails exclusively for the property defined in
Section~\ref{sec:sldsemantics} as the negation of \succeeds, which also
includes the case of \nontermination. Consequently, the names
\terminates and \succeeds in \lptp documents refer to the same
properties described under those names in Section~\ref{sec:sldsemantics}.

The second set of notions is the most important in practice, as it is
the one handled directly by the \lptp system. In the system, we have
access to three syntactic operators acting on program goals, transforming
them into formulas in the formal language of \lptp. Formally, we write
$\textbf{S}$, $\textbf{F}$, $\textbf{T}$ for the succeed, failure and
termination operators, respectively. The complete inductive definition
can be found in \cite{LPTP-Stark97} along with the following theorems
that relate the two notions:

\exabeg
\begin{theorem}[Soundness]
~
\vspace*{-0.5mm}  
\begin{enumerate}
  \itemsep=0pt
\item If $G$ \terminates, then $\text{IND}(P) \vdash \mathbf{T}\,G$.
\item If $G$ \succeeds with answer $\sigma$, then $\text{IND}(P) \vdash \mathbf{S}\,G\sigma$.
\item If $G$ \finitelyfailstext, then $\text{IND}(P) \vdash \mathbf{F}\,G$.
\end{enumerate}

\end{theorem}
\exaend

%
\begin{theorem}[Adequacy]
~
\vspace*{-1mm}  
\begin{enumerate}
  \itemsep=0pt
\item If $\text{IND}(P) \vdash \mathbf{T}\,G$, then $G$ \terminates.
\item If $\text{IND}(P) \vdash \mathbf{T}\,G \land \mathbf{S}\,G\sigma$, then $G$ \succeeds with answer including $\sigma$.
\item If $\text{IND}(P) \vdash \mathbf{T}\,G \land \mathbf{F}\,G$, then $G$ \finitelyfailstext.
\end{enumerate}

\end{theorem}
\exaend

\noindent
Note that for a terminating goal we get equivalence between the two notions of success and failure but not in general.

\secbeg
\section{SLD Semantics and \lptp} \label{sec:sldsemanticsandlptp}
\secend

In this section we discuss the relation between the properties
declaratively defined in Section~\ref{sec:sldsemantics} and the \lptp
operators.
We first relate the properties
that have been defined with respect to a breadth-first search strategy
that generates the complete SLD-tree
(Section~\ref{sec:prop-trees-breadth}), i.e., universal properties.
Then, we carry out the same analysis for properties defined with
respect to \prolog’s depth-first search strategy
(Section~\ref{sec:prop-trees-depth}). 

Both kinds of properties coincide for terminating goals.

However, for non-terminating goals (i.e., goals for which there
exist infinite branches), differences arise. These appear when the
definition of the property in question may involve both \infinite and
\successful branches of the search trees. In such cases, the tree
reachable under depth-first search before looping may contain no
successful branch, whereas the breadth-first tree may contain
successful branches.
The resulting relationships are summarized in
Tables~\ref{tab:bf-props-and-lptp-correspondence},
~\ref{tab:bf-df-prop-correspondence} and 
~\ref{tab:ciaopp-props-and-lptp-correspondence}.

\begin{table}[t]
\centering
\vspace*{-3mm}
\begin{tabular}{lll}
\hline
\textbf{BF (Universal) Property} 
& \textbf{Relation} 
& \textbf{\lptp Inference} \\
\hline

$\terminates(P,Q)$
& $\Leftrightarrow$
& $\text{IND}(P) \vdash \mathbf{T}\,Q$ \\

$\noterminates(P,Q)$
& $\Leftrightarrow$
& $\text{IND}(P) \vdash \neg \mathbf{T}\,Q$ \\

$\succeeds(P,Q)$
& $\Leftrightarrow$
& $\text{IND}(P) \vdash \mathbf{S}\,Q$ \\

$\fails(P,Q)$
& $\Leftrightarrow$
& $\text{IND}(P) \vdash \mathbf{F}\,Q$ \\

$\finitelysucceeds(P,Q)$
& $\Leftrightarrow$
& $\text{IND}(P) \vdash \mathbf{T}\,Q \land \mathbf{S}\,Q$ \\

$\infinitelysucceeds(P,Q)$
& $\Leftrightarrow$
& $\text{IND}(P) \vdash \neg \mathbf{T}\,Q \land \mathbf{S}\,Q$ \\

$\finitelyfails(P,Q)$
& $\Leftrightarrow$
& $\text{IND}(P) \vdash \mathbf{T}\,Q \land \mathbf{F}\,Q$ \\

$\infinitelyfails(P,Q)$
& $\Leftrightarrow$
& $\text{IND}(P) \vdash \neg \mathbf{T}\,Q \land \mathbf{F}\,Q$ \\

$not\_finitely\_fails(P,Q)$
& $\Leftrightarrow$
& $\text{IND}(P) \vdash \mathbf{S}\,Q \lor \neg \mathbf{T}\,Q$ \\

\hline
\end{tabular}
\vspace*{-2mm}
\caption{Relating universal properties (breadth-first) with properties inferred by \lptp.} %
\label{tab:bf-props-and-lptp-correspondence}
\vspace*{-2mm}
\end{table}

\begin{table}[t]
\centering
\begin{tabular}{lll}
\hline
\textbf{DF (\prolog) Property} 
& \textbf{Relation} 
& \textbf{BF (Universal) Property} \\
\hline

$\terminates_{df}(P,Q)$
& $\Leftrightarrow$
& $\terminates(P,Q)$
\\

$\noterminates_{df}(P,Q)$
& $\Leftrightarrow$
& $\noterminates(P,Q)$
\\

$\succeeds_{df}(P,Q)$
& $\Rightarrow$
& $\succeeds(P,Q)$
\\

$\fails_{df}(P,Q)$
& $\Leftarrow$
& $\fails(P,Q)$
\\

$\finitelysucceeds_{df}(P,Q)$
& $\Leftrightarrow$
& $\finitelysucceeds(P,Q)$
\\

$\infinitelysucceeds_{df}(P,Q)$
& $\Rightarrow$
& $\infinitelysucceeds(P,Q)$
\\

$\finitelyfails_{df}(P,Q)$
& $\Leftrightarrow$
& $\finitelyfails(P,Q)$
\\

$\infinitelyfails_{df}(P,Q)$
& $\Leftarrow$
& $\infinitelyfails(P,Q)$
\\

$not\_finitely\_fails_{df}(P,Q)$
& $\Leftrightarrow$
& $not\_finitely\_fails(P,Q)$
\\

\hline
\end{tabular}
\vspace*{-2mm}
\caption{Relating DF (\prolog) Properties with BF (Universal)
  Properties.}
\label{tab:bf-df-prop-correspondence}
\vspace*{-1mm}
\end{table}

\begin{table}[t]
\centering
\begin{tabular}{lll}
\hline
\textbf{\lptp Inference} 
& \textbf{Relation} 
& \textbf{\ciaopp Inference} \\
\hline
$\text{IND}(P) \vdash \mathbf{T}\,G$
& $\Leftrightarrow$
& $\terminates_{ciaopp}(P,Q)$ \\
$\text{IND}(P) \vdash \neg \mathbf{T}\,G$
& $\Leftrightarrow$
& $\noterminates_{ciaopp}(P,Q)$ \\
$\text{IND}(P) \vdash \mathbf{S}\,G$ & $\Leftrightarrow$ & $\succeeds_{ciaopp}(P,Q)$ \\
              & $\Leftarrow$ & $not\_finitely\_fails_{ciaopp}(P,Q) \land \terminates_{ciaopp}(P,Q)$ \\
$\text{IND}(P) \vdash \mathbf{F}\,G$
& $\Leftrightarrow$
& $\fails_{ciaopp}(P,Q)$ \\
$\text{IND}(P) \vdash \mathbf{T}\,G \land \mathbf{S}\,G$
& $\Leftrightarrow$
& $\finitelysucceeds_{ciaopp}(P,Q)$ \\
& $\Leftrightarrow$ & $not\_finitely\_fails_{ciaopp}(P,Q) \land \terminates_{ciaopp}(P,Q)$ \\
$\text{IND}(P) \vdash \neg \mathbf{T}\,G \land \mathbf{S}\,G$
& $\Leftrightarrow$
& $\infinitelysucceeds_{ciaopp}(P,Q)$
\\
$\text{IND}(P) \vdash \mathbf{T}\,G \land \mathbf{F}\,G$
& $\Leftrightarrow$
& $\finitelyfails_{ciaopp}(P,Q)$
\\
$\text{IND}(P) \vdash \neg \mathbf{T}\,G \land \mathbf{F}\,G$
& $\Leftrightarrow$
& $\infinitelyfails_{ciaopp}(P,Q)$
\\
$\text{IND}(P) \vdash \mathbf{S}\,G \lor \neg \mathbf{T}\,G$
& $\Leftrightarrow$
& $not\_finitely\_fails_{ciaopp}(P,Q)$ \\

\hline
\end{tabular}
\vspace*{-2mm}
\caption{Relating properties inferred by \lptp and properties inferred by \ciaopp.}
\label{tab:ciaopp-props-and-lptp-correspondence}
\vspace*{-2mm}
\end{table}

Table~\ref{tab:ciaopp-props-and-lptp-correspondence} expresses
relations between properties inferred by \lptp and properties inferred
by \ciaopp. They allow the bidirectional interaction between the two
complementary systems. For example, if a property is inferred by
\ciaopp, using such relations, we could assert a logical
representation of such property as an axiom for \lptp or as a proof
with an special tag ``\texttt{by \ciaopp}''. Conversely, properties
inferred by \lptp could be expressed in some cases as \ciao assertions
with status \texttt{trust}.

This concludes the formalization of the semantic framework. Building
on these correspondences, we now address the practical translation of
\ciao assertions into \lptp theorems.

\secbeg
\section{From \ciao Assertions to Logical Formulae}
\secend
\label{sec:ciao_to_logic}

First, we propose a suitable representation of \ciao
assertions as first-order logical formulas. We focus for now on
\ciao assertion schemas of the form:

\prettylstciao
\begin{lstlisting}
:- pred p(X1, ..., Xn) : Pre => Post + Comp.
\end{lstlisting}
\prettylstciao

\noindent
which, as described in Section~\ref{sec:background}, are internally
translated into an equivalent set of three assertions:

\prettylstciao
\begin{lstlisting}
:- calls p(X1, ..., Xn) : Pre.
:- success p(X1, ..., Xn) : Pre => Post.
:- comp p(X1, ..., Xn) : Pre + Comp.
\end{lstlisting}
\prettylstciao

The \texttt{calls} assertion presents problems due to its essentially operational
nature but both \texttt{success} and \texttt{comp} assertions can be
translated to first order formulas which are easily interpreted in the
\lptp language.
The properties in the \texttt{Pre} and \texttt{Post} fields are built from
conjunctions of predicates over the variables of \texttt{p/n}, or from
\emph{star terms}, a shorthand notation that ultimately reduces to
such conjunctions. Because of this, they can be expressed directly in
the language of $P$.
The situation is different for \texttt{Comp}. Its properties describe
the execution behavior of a program, which falls outside what the base language of
$P$ can express. Nevertheless, many of these properties can be
captured in the extended language $\text{IND}(P)$.

We will later make explicit this
representation, for now we assume that we are working in an
appropriate extension of the language in which we can express a
formula $\mathsf{Comp}$ representing this property.

Finally, we can represent the \texttt{success} assertion via the formula: 
\vspace*{-3mm}
\begin{equation}\label{assert:succeeds}
  \forall \bar{X}: \mathsf{Pre}(\bar{X}) \rightarrow (\mathsf{succeeds}(p(\bar{X})) \rightarrow \mathsf{Post}(\bar{X}))
\end{equation}
\vspace*{-7mm}

\noindent
where $\bar{X} = (X_1, \dots, X_n)$ is an $n$-tuple of variables and
$\mathsf{succeeds}$ is the \lptp operator. And the \texttt{comp}
assertion via:

\vspace*{-8mm}
\begin{equation}\label{assert:comp}
    \forall \bar{X}: \mathsf{Pre}(\bar{X}) \rightarrow \mathsf{Comp}(p(\bar{X})).
\end{equation}

\vspace*{-2 mm}
Notice that both expressions are of the form:
\vspace*{-3mm}
\begin{equation}\label{assert:general}
  \forall \bar{X}: \mathsf{Pre}(\bar{X}) \rightarrow \Psi(\bar{X}).
\end{equation}
\vspace*{-8mm}

\noindent where $\Psi$ is a predicate in the adequate language. Moreover, we can put both formulas
together into a single formula for a direct translation of the original \texttt{pred} assertion:

\vspace*{-3mm}
\begin{equation}\label{assert:comb}
  \forall \bar{X}: \mathsf{Pre}(\bar{X}) \rightarrow ((\mathsf{succeeds}(p(\bar{X})) \rightarrow \mathsf{Post}(\bar{X})) \land \mathsf{Comp}(p(\bar{X}))).
\end{equation}
\vspace*{-6mm}

We now apply the general translation scheme (Eq. 5) to concrete \ciao
properties. The sections below demonstrate how both simple properties
(types) and complex computational properties (termination,
non-failure) fit within our framework, revealing which assertions are
naturally encodable and which require approximation.

\secbeg
\section{Property Correspondence}
\label{sec:prop_corresp}
\secend

As previously explained, both \ciao and \lptp assertions are formed from
a collection of atomic predicate symbols that state properties about
the program's predicates. In \lptp these atomic properties are three:
$\mathsf{succeeds}$, $\mathsf{fails}$ and $\mathsf{terminates}$. In \ciao
we have more complex atomic properties such as determinacy or not
finitely failing but a lot of them can be represented in \lptp using its three
atomic properties and its complete assertion language. In this section
we provide a correspondence between \ciao properties and \lptp
assertions.
Table~\ref{tab:prop-correspondence} summarizes the correspondence
between the \ciao properties used in assertions and their
representation in \lptp. Throughout this section we restrict our
attention to \emph{instantiation types}.

\begin{table}[t]
  \centering
  \small
  \renewcommand{\arraystretch}{1.15}
  \begin{tabular}{|>{\centering\arraybackslash}m{3.5cm}|m{9.5cm}|}
    \hline
    \textbf{\ciao property} & \textbf{Corresponding \lptp formula} \\
    \hline
    Type atom \(q(V)\) in \(\mathbf{Pre}\) or \(\mathbf{Post}\)
    & \(\lptpS \mathsf{q}(V)\) \\
    \hline
    \texttt{ground(V)}
    & \(\mathsf{gr}(V)\) \\
    \hline
    \texttt{terminates}
    & \(\lptpT\,\mathsf{p}(X)\) \\
    \hline
    \texttt{not\_fails}
    & \(\exists X_2:\ \neg \lptpT\,\mathsf{p}(X_1,X_2)\ \lor\ \lptpS\,\mathsf{p}(X_1,X_2)\) \\
    \hline
    \texttt{num\_solutions(X,N)}
    & \(\exists_{=N} x:\ \mathsf{p}(x)\) \\
    \hline
    \texttt{det(X)}
    & \(\exists_{=N} (x, t):\ \mathsf{p}^{t}(x,t)\), using an explicit trace argument; equivalently, one successful trace or divergence \\
    \hline
  \end{tabular}
  \vspace*{-2mm}
  \caption{Correspondence between \ciao properties and \lptp formulas.}
  \label{tab:prop-correspondence}
\end{table}

\smallskip
\paragraph{Type and Mode Properties.}
The first two rows of Table~\ref{tab:prop-correspondence} are direct
translations. A type property appearing in the \(\mathbf{Pre}\) or
\(\mathbf{Post}\) part of a \ciao assertion is represented in \lptp\ by
a \(\mathsf{succeeds}\) atom for the corresponding predicate, while
\texttt{ground/1} is translated as \(\mathsf{gr/1}\).
For example, consider the following program:\\

\prettylstciao
\begin{lstlisting}
nat(0).
nat(s(X)) :- nat(X).

plus(0,Y,Y).
plus(s(X),Y,s(Z)) :- plus(X,Y,Z).

times(0,_Y,0).
times(s(X),Y,Z) :- times(X,Y,A), plus(Y,A,Z).
\end{lstlisting}
\prettylstciao

The following \ciao assertion:
\prettylstciao
\begin{lstlisting}
:- success plus(X, Y, Z) : (nat(X), nat(Y)) => nat(Z).
\end{lstlisting}
\prettylstciao
\noindent
expresses that if any call to plus(X, Y, Z) with X and Y bound to
natural numbers (nat) succeeds, then Z must be bound to a nat upon success. It
corresponds, by the type row of Table~\ref{tab:prop-correspondence},
to:
$\forall (X,Y,Z).\;
  \left(
    \lptpS \mathsf{plus}(X,Y,Z)
    \land \lptpS \mathsf{nat}(X)
    \land \lptpS \mathsf{nat}(Y)
  \right)
  \rightarrow
  \lptpS \mathsf{nat}(Z).$

\noindent
In \lptp\ this can be stated as:
\prettylstciao
\begin{lstlisting}
:- lemma(plus:type,
all [x,y,z]: succeeds plus(?x,?y,?z) & succeeds nat(?x) & succeeds nat(?y) =>
 succeeds nat(?z), [...]).
\end{lstlisting}
\prettylstciao
\noindent
This assertion can be proven automatically by \ciaopp %
provided we %
add an extra assertion: %
\prettylstciao
\begin{lstlisting}
:- entry plus(X, Y, Z) : (nat(X), nat(Y)).
\end{lstlisting}
\prettylstciao

\noindent
This explicitly tells the preprocessor about the way this predicate is
going to be called, which
is necessary due to the top-down and goal-driven nature of the \ciaopp
system.

The same direct translation applies to mode properties. For instance,
the following \ciao assertion:
\prettylstciao
\begin{lstlisting}
:- success plus(X, Y, Z) : (ground(X), ground(Y)) => ground(Z).
\end{lstlisting}
\prettylstciao
corresponds, by the groundness row of Table~\ref{tab:prop-correspondence}, to:
\prettylstciao
\begin{lstlisting}
:- lemma(plus:ground,
all [x,y,z]: succeeds plus(?x,?y,?z) & gr(?x) & gr(?y) => gr(?z), [...]).
\end{lstlisting}
\prettylstciao

It is also worth noting why the \(\mathsf{succeeds}\) atom appears in
the antecedent of the translation. If we replace \(\mathsf{plus}\) by
the failing predicate:
\prettylstciao
\begin{lstlisting}
plusB(0,Y,Y) :- fail.
plusB(s(X),Y,s(Z)) :- plus(X,Y,Z).
\end{lstlisting}
\prettylstciao
then the translated implication may still hold vacuously, since \(\mathsf{plusB}\) never succeeds. Indeed, the following lemma is provable in \lptp:

\prettylstciao
\begin{lstlisting}
:- lemma(plusB:fail, all [x,y,z]: succeeds nat(?x) => fails plusB(?x,?y,?z), [...]).
\end{lstlisting}
\prettylstciao

\paragraph{Computational Properties.}
The remaining rows of Table~\ref{tab:prop-correspondence} concern
computational properties. For termination, the translation is immediate:
a \ciao assertion of the form: 
\prettylstciao
\begin{lstlisting}
:- comp plus(X, Y, Z) : (nat(X), nat(Y)) + terminates.
\end{lstlisting}
\prettylstciao
becomes:
\prettylstciao
\begin{lstlisting}
:- lemma(plus:termination:1,
all [x,y,z]: succeeds nat(?x) => terminates plus(?x,?y,?z), [...]).
\end{lstlisting}
\prettylstciao
as indicated by the \texttt{terminates} row.
For \texttt{not\_fails}, the corresponding row in
Table~\ref{tab:prop-correspondence} expresses that a call either
succeeds for some output or does not terminate. Writing
\(X=(X_1,X_2)\), with \(X_1\) the input variables and \(X_2\) the
output variables, the following \ciao assertion:
\prettylstciao
\begin{lstlisting}
:- comp plus(X, Y, Z) : (nat(X), nat(Y)) + not_fails.
\end{lstlisting}
\prettylstciao
is translated as:
 \prettylstciao
\begin{lstlisting}
:- lemma(plus:not_fails,
all [x,y]: succeeds nat(?x) =>
  (ex z: ~ terminates plus(?x,?y,?z) \textbackslash / succeeds plus(?x,?y,?z)), [...]).
\end{lstlisting}
\prettylstciao

For \texttt{num\_solutions(X,N)}, Table~\ref{tab:prop-correspondence}
uses the counting quantifier \(\exists_{=N}\). Since \lptp\ is first-order,
this should be understood as syntactic sugar for a first-order formula
expressing that there are exactly \(N\) distinct solutions. In the
special case \(N=1\), this yields the usual uniqueness statement; for
example, uniqueness of Peano addition can be written as:
\prettylstciao
\begin{lstlisting}
:- lemma(plus:uniqueness,
all [x,y,z1,z2]:
  succeeds plus(?x,?y,?z1) & succeeds plus(?x,?y,?z2) => ?z1 = ?z2, [...]).
\end{lstlisting}
\prettylstciao

Finally, the last row of Table~\ref{tab:prop-correspondence} explains
\texttt{det(X)}. Since \ciao determinacy distinguishes solutions not
only by answer substitutions but also by derivation trace, the
translation is not stated directly on the original predicate; instead,
we use a traced version \(\mathsf{p}^{t}(X,t)\) with an additional
trace argument and then require exactly one successful trace.
For example, consider:
\prettylstciao
\begin{lstlisting}
edge(a, b).    edge(b, d).    edge(a, c).    edge(c, d).

path(X, Y) :- edge(X, Y).
path(X, Y) :- edge(X, Z), path(Z, Y).
\end{lstlisting}
\prettylstciao
The query \lstinline|path(a,d)| yields the same output binding along two different derivations, so to capture determinacy we introduce traces explicitly:
\prettylstciao
\begin{lstlisting}
path_trace(X,Y, c1(T)) :- edge(X,Y, T).
path_trace(X,Y, c2(T1,T2)) :- edge(X,Z,T1), path_trace(Z,Y, T2).
\end{lstlisting}
\prettylstciao
Determinacy is then expressed %
applying the \texttt{num\_solutions} translation with \(N=1\) to the traced predicate.

Finally, using the property correspondence between \ciao and \lptp 
above, and the translation schemas given in
Section~\ref{sec:ciao_to_logic}, we conclude that if we can find a
proof for the proposed translation
of a \ciao assertion into
\lptp, then such assertion is \emph{true} (w.r.t.\ 
definitions given in Section~\ref{sec:assertion-checking}).
Conversely, \emph{true} assertions inferred by \ciaopp could
be stated as theorems in \lptp.

\section{Conclusions} \label{sec:conclusions}
\secend

We have addressed the problem of relating Abstract
Interpretation–based verification and Theorem Proving by studying the
translation of \ciao assertions into \lptp formulae and identifying a
partial correspondence between assertion-based and logic-based
specifications, showing that while many assertions translate
faithfully, others resist direct encoding due to intrinsic semantic
mismatches.  We have introduced a partial translation scheme,
characterized assertion classes according to their logical
encodability, and proposed approximation strategies and auxiliary
constructs for non-translatable cases. We have also analyzed the
resulting soundness and completeness trade-offs. We argue that the
proposed techniques enables a tight integration of \ciao’s assertion
checking with \lptp-based deductive verification, leveraging their
complementary capabilities.

\clearpage
\bibliographystyle{eptcs}
\bibliography{references_used}
\end{document}